# Influence of deposition parameters on the plasmonic properties of gold nanoantennas fabricated by focused ion beam lithography

Michael Foltýn[1], Marek Patočka[1,2], Rostislav Řepa[1], Tomáš Šikola[1,3], Michal Horák[3,*]

[1] Brno University of Technology, Faculty of Mechanical Engineering, Institute of Physical Engineering, Technická 2, 616 69, Brno, Czech Republic

[2] NenoVision, Purkyňova 127, 612 00, Brno, Czech Republic

[3] Brno University of Technology, Central European Institute of Technology, Purkyňova 123, 612 00, Brno, Czech Republic

[*] michal.horak2@ceitec.vutbr.cz

**Abstract**

The behavior of plasmonic antennas is influenced by a variety of factors, including their size, shape, and material. Even minor changes in the deposition parameters during the thin film preparation process may have a significant impact on the dielectric function of the film, and thus on the plasmonic properties of the resulting antenna. In this work, we deposited gold thin films with thicknesses of 20 nm, 30 nm, and 40 nm at various deposition rates using an ion-beam-assisted deposition. We evaluate their morphology and crystallography by atomic force microscopy, X-ray diffraction, and transmission electron microscopy. Next, we examined the ease of fabricating plasmonic antennas using focused-ion-beam lithography. Finally, we evaluate their plasmonic properties by electron energy loss spectroscopy measurements of individual antennas. Our results show that the optimal gold thin film for plasmonic antenna fabrication of a thickness of 20 and 30 nm should be deposited at the deposition rate of around 0.1 nm/s. The thicker 40 nm film should be deposited at a higher deposition rate like 0.3 nm/s.

**Introduction**

Localized surface plasmon resonances (LSPRs) are standing waves of an electromagnetic field related to electron gas oscillations at a metal-dielectric interface. LSPRs are known for their ability to be confined into a space at the subwavelength scale[1,2], giving multiple possibilities for applications[3] including biosensing[4,5] and use in ultrathin optical systems[6,7]. Gold has been the material of choice for plasmonics for many years due to its chemical stability, biological compatibility[8], and its relative ease of preparation[9,10]. Properties of LSPRs depend on the dielectric function of the used material, which is affected by crystallinity, size of grains, grain boundaries, and surface roughness of the material[11-15], and, consequently, by fabrication methods, including top-down methods[16]. In the case of polycrystalline thin films, the final morphology is given by the initial states of layer growth[17,18] and may be tuned by choosing a proper substrate and by optimizing deposition parameters such as temperature, operation pressure, and deposition rate.

The substrate has a major influence on the morphology of the resulting thin films. A smooth substrate surface is essential for creating a uniform boundary between the substrate and the deposited layer. This improves the film adhesion and reduces the occurrence of defects and voids[17,19,20]. In some cases, a thin adhesion layer is deposited first to improve thin film adhesion. While this has a favorable impact on the resultant film structure and adhesion, it harms the plasmonic properties of the antennas[21,22]. The temperature influences the size of grains. An increased temperature of both the substrate and the target material during the deposition leads to larger grains of the resulting film[23-25] or larger nanoparticles in the case of liquid substrates[24-26]. The pressure inside the deposition chamber influences the size of grains and nanoparticles and their crystallographic orientation. A higher pressure results in larger grains and nanoparticles[24,27] and a reduction of the crystallographic orientation diversity[28]. A uniform crystallographic composition is advantageous, as it limits the preferential milling during the focused ion beam lithography[15,29,30]. The deposition rate is one of the most used methods for influencing the properties of deposited films. For ultrathin films, higher deposition rates have either little or even no impact on the structure of the films. However, in the case of thicker films, the deposition rate becomes a crucial factor. A higher deposition rate results in a smoother surface and fewer defects and sometimes facilitates the formation of larger grains[31-33]. Both larger grains and reduced surface roughness result in lower energy losses of LSPRs, thus improving their plasmonic properties[34-36]. The grain size can also be increased by annealing after deposition[36-38]. Consequently, to obtain optimal thin films for plasmonic applications, it is essential to use smooth substrates, high deposition rates, higher deposition temperatures, and higher pressure. Despite many published works dealing with the properties of plasmonic antennas, there is no experimental paper discussing the impact of experimental growth conditions on the plasmonic properties of resulting metallic antennas.

In our contribution, we evaluate how the deposition rate of gold thin films affects the fabrication yield and plasmonic properties of antennas made by focused ion beam (FIB) lithography. Gold polycrystalline thin films of a thickness of 20 nm, 30 mm, and 40 nm were deposited on standard silicon nitride membranes for transmission electron microscopy at four different deposition rates reading 0.2 Ås$^{-1}$, 1 Ås$^{-1}$, 2 Ås$^{-1}$, and 3 Ås$^{-1}$ in a custom-built deposition chamber (Figure 1A)[39]. We have studied the structural properties of the films by atomic force microscopy (AFM), X-ray diffraction (XRD), selective area electron diffraction (SAED), and scanning transmission electron microscopy (STEM) using the annular dark field (ADF) and high-angle annular dark field (HAADF) detector. The films were further processed by FIB to prepare the plasmonic antennas (Figure 1B)[15,16]. To evaluate the influence on the fabrication yield and plasmonic properties of the antennas, we fabricated three distinct antenna types (Figure 1 C): narrow and wide bar antennas with dimensions of 240 × 40 nm$^2$ and 240 × 80 nm$^2$, respectively, and bow-tie antennas with a total length of 500 nm. Plasmonic properties have been studied by STEM combined with electron energy loss spectroscopy (EELS) individual structures[40-42]. We note that the wider bar antennas have been already chosen as the test structures for our previous comparative study of polycrystalline and monocrystalline antennas[15]. Similarly, the bow-tie antennas have been studied in our group both theoretically[43] and experimentally[44]. Hence, these types of antennas have been taken for this study as well, as they represent a well-known system.

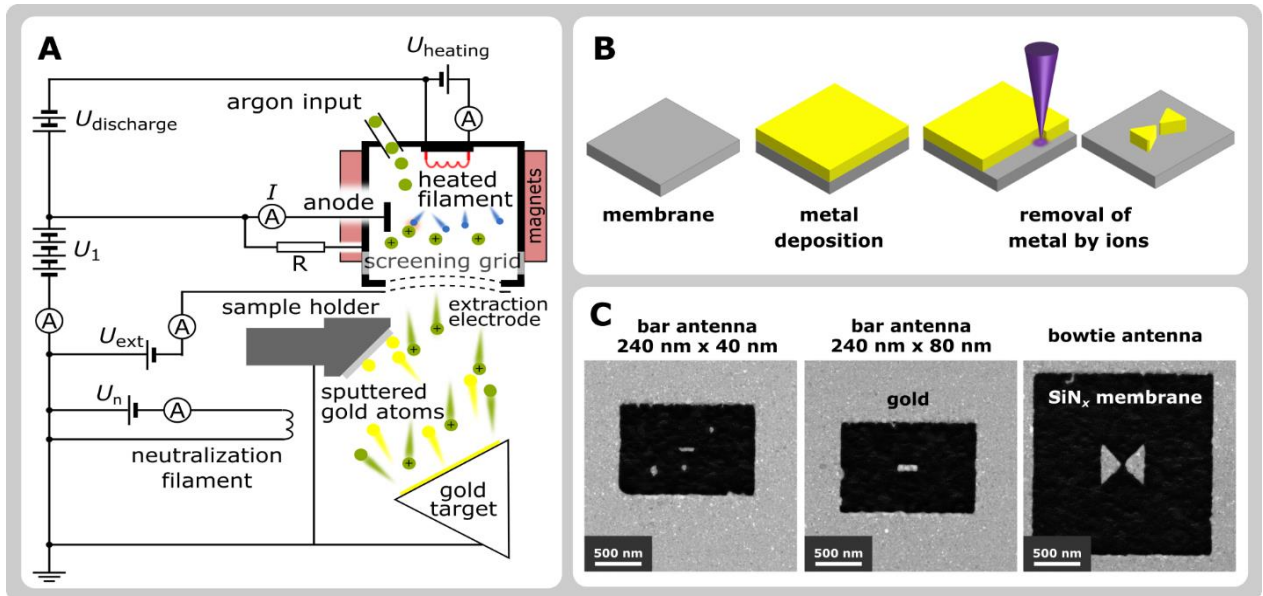

*Figure 1. Fabrication of plasmonic antennas. **A**: Schematic of a deposition chamber with the Kaufman broad ion beam source. Argon working gas is ionized by electrons thermally emitted from the heated filament and confined by a magnetic field inside the discharge chamber. These ions are extracted by the extraction voltage $U_{ext}$ and accelerated by the acceleration voltage $U_1$. The sputter ion yield and thus the thin film deposition rate is then given by the ion beam current and energy being controlled by the discharge current I and acceleration voltage $U_1$. **B**: Schematic of plasmonic antenna fabrication by FIB lithography. **C**: STEM-HAADF micrographs of three types of fabricated antennas. The grey color represents the antennas and the remaining gold, while the black color corresponds to the area where the gold layer was removed by FIB.*

**Experimental details**

Gold was deposited by ion beam sputtering of a gold target (Kurt J. Lesker Company) under normal ion beam incidence on standard 30-nm-thick silicon nitride membranes for TEM with a window size of 250 × 250 μm$^2$ and frame thickness of 200 μm by Agar Scientific in a custom-built deposition chamber utilizing the Kaufman broad ion beam source (Figure 1A)[39]. The deposition pressure was in the order of $10^{-3}$ Pa and the deposition rates were 0.2 Ås$^{-1}$, 1 Ås$^{-1}$, 2 Ås$^{-1}$, and 3 Ås$^{-1}$. The deposition rate was controlled by the argon flux, filament current, and discharge voltage (all of them determining the discharge current), and by the acceleration voltage. During the deposition, the thickness of the film and the deposition rate were measured in situ by a quartz crystal microbalance monitor.

AFM measurements were performed using the LiteScope microscope (NenoVision) with the Akiyama self-sensing probe (resonant frequency ~ 45 kHz, spring constant ~ 5 N/m, tip radius < 15 nm) in the frequency-modulated tapping regime under ambient conditions. The measured area was 4 µm$^2$, scanning window 2 × 2 µm$^2$ (512 × 512 pixels with a pixel size of 3.9 × 3.9 nm$^2$), and scanning speed 1.5 µm/s. For each scanned sample, two positions in the vicinity of the SiN$_x$ membrane approximately 120 µm apart were randomly selected to perform scans.

XRD measurement was carried out by the Rigaku SmartLab 3 kW powder diffractometer in the parallel beam geometry with added divergence slit, anti-scatter slit, and Soller slits. The diffractograms were analyzed by Rietveld analysis in Profex software.

FIB lithography was done using the dual beam FIB/SEM microscope FEI Helios with gallium ion beam with an energy of 30 keV and an ion beam current of 1.7 pA. We note that the energy (the highest available) and the current (the lowest available) are optimized for the best spatial resolution of the milling.

Transmission electron microscopy was performed with TEM FEI Titan equipped with a GIF Quantum spectrometer. SAED and STEM measurements were done at 300 kV. SAED patterns were recorded out of the area of 9.1 µm$^2$. STEM-EELS measurements were performed in a monochromated scanning regime at 120 kV. The beam current was set to 0.2 nA and the FWHM of the zero-loss peak was around 0.10 eV. We set the convergence angle to 10 mrad, the collection angle to 11.4 mrad, and the dispersion of the spectrometer to 0.01 eV/pixel. These are the optimal parameters to measure LSPRs by STEM-EELS[45].

**Results**

**Structure and morphology of thin films**

Gold layers with thicknesses of 20 nm and 30 nm were deposited at deposition rates reading 0.2 Ås$^{-1}$, 1 Ås$^{-1}$, and 3 Ås$^{-1}$. The 40-nm-thick film was grown at higher deposition rates reading 1 Ås$^{-1}$, 2 Ås$^{-1}$, and 3 Ås$^{-1}$. The morphology and structural parameters of 30-nm-thick gold films studied using AFM, XRD, and STEM are shown in Figure 2.

Figure 2A depicts the morphology of thin films grown at deposition rates of 0.2 Ås$^{-1}$, 1 Ås$^{-1}$, and 3 Ås$^{-1}$. One can see that with the increasing deposition rate, the island-like structure starts to be more profound. The thin film deposited at 0.2 Ås$^{-1}$ is generally smooth and uniform, punctuated with a few protrusions. The film deposited at 1 Ås$^{-1}$ is uniformly covered with small spherical objects – islands. Finally, the film deposited at 3 Ås$^{-1}$ shows out even bigger islands. Consequently, the root mean square (RMS) roughness evaluated from the AFM micrographs increases with the deposition rate reading 0.5 nm, 1.4 nm, and 3.8 nm, respectively.

The crystallography analysis of the films done by XRD provided the information on average size of grains and their preferential crystallographic orientations perpendicular to the surface. The measured diffractograms showing considerable similarities for all deposition rates are shown in Figure 2B. They contain an intense peak at 38° corresponding to the (111) reflection and a much lower peak at 43° corresponding to the (200) reflection. Moreover, the diffractogram for the deposition rate of 0.2 Ås$^{-1}$ contains a peak at 81° which is slightly above the noise level and corresponds to the (222) reflection. Consequently, the dominant crystallographic orientation is the (111) plane for all three films. The average crystal grain size was evaluated from the (111) reflection peak using the Debye-Scherrer equation. With the Scherrer constant chosen as 0.9[46], the average (111) grain size reads (14.97 ± 0.15) nm, (14.77 ± 0.13) nm, and (12.68 ± 0.17) nm for the films deposited at 0.2 Ås$^{-1}$, 1 Ås$^{-1}$, and 3 Ås$^{-1}$, respectively. Consequently, the average size of the (111) grains reduces with the increasing deposition rate. In the case of (200) and (222) reflection peaks, such a quantitative analysis of the grain size was impossible due to a low signal-to-background ratio. Both the preferential orientation and measured grain size for the (111) orientation are in agreement with similar XRD studies of sputtered gold thin films[47,48].

The crystallography of these thin films was locally analyzed by transmission electron microscopy using SAED and STEM to support the XRD results. Obtained diffraction patterns (Figure 2C) show not only the crystallographic orientations of film grains identified by XRD but also additional ones, which were below the detection limit of XRD. Moreover, STEM-ADF micrographs of the films (Figure 2D) show that the film deposited at 0.2 Ås$^{-1}$ has less pronounced boundaries between grains compared to the other two films. Our STEM measurements also verify the grain sizes obtained by XRD, with the film deposited at 3 Ås$^{-1}$ having smaller grains than the other two films.

Based on the structural analysis of 30 nm films, the films deposited at 0.2 Ås$^{-1}$ and 1 Ås$^{-1}$ appear to be ideal candidates for use in plasmonics. Both have larger grains and lower surface roughness compared to that deposited at 3 Ås$^{-1}$. Unfortunately, the less pronounced grain boundaries present in the film deposited at 0.2 Ås$^{-1}$ might complicate the fabrication of plasmonic antennas, leading to inclined edges of structures fabricated by FIB lithography[15].

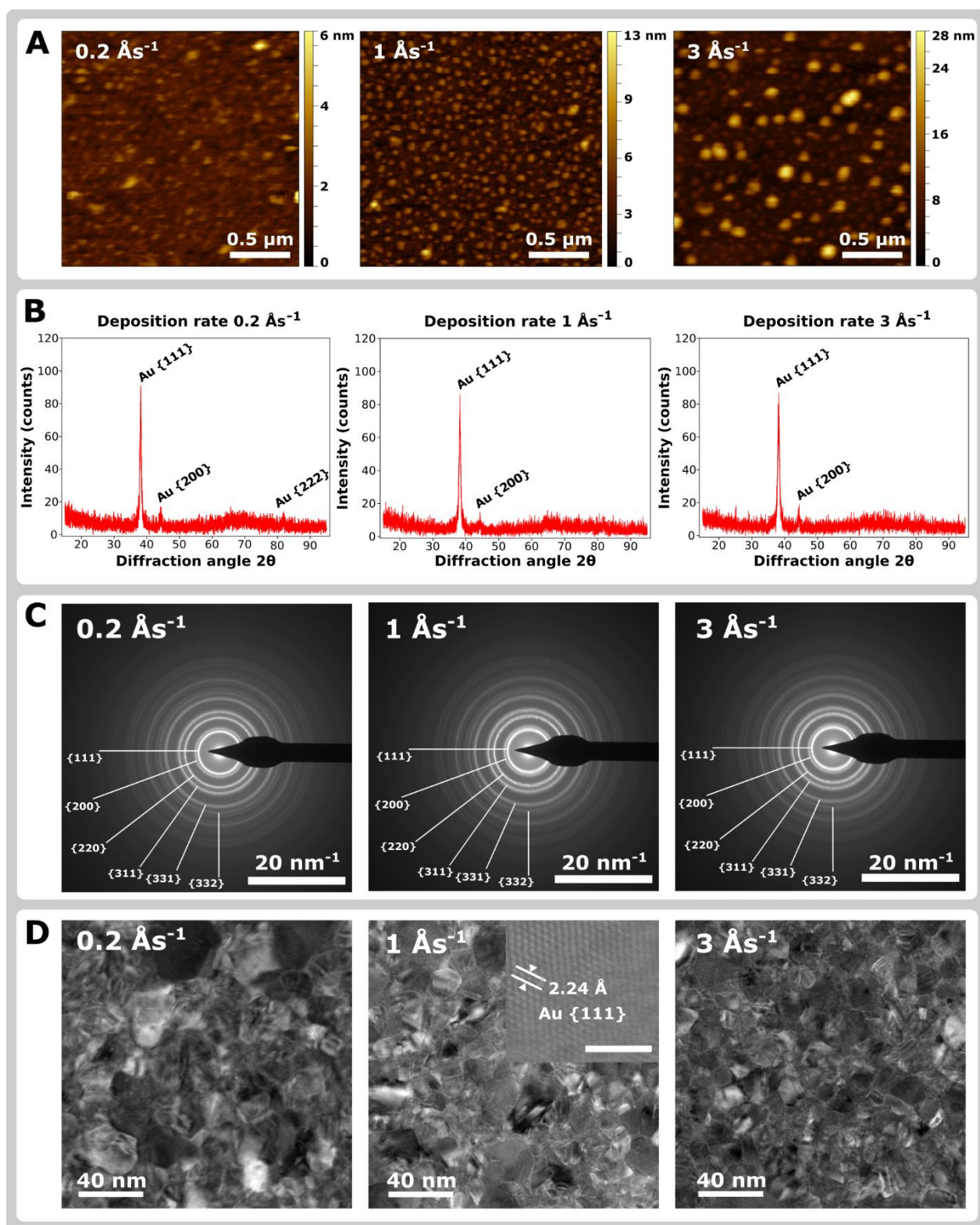

*Figure 2*. Structure and morphology of 30-nm-thick gold films deposited at 0.2 Ås$^{-1}$, 1 Ås$^{-1}$, and 3 Ås$^{-1}$. **A:** Surface profile measured by AFM. **B:** X-ray diffractograms. **C:** SAED patterns. **D:** STEM-ADF micrographs with an inset showing the STEM-HAADF micrograph of the Au{111} lattice (the scale bar in the inset is 2 nm long).

**Fabrication yield**

The fabrication yield is a crucial indicator for the evaluation of the efficiency of plasmonic antenna fabrication by FIB lithography. It is calculated as the total number of proper antennas divided by the total number of fabricated antennas. The higher the fabrication yield, the better the film is for fabrication of plasmonic antennas. Antennas considered as the proper ones are free of residual grains and retain their original shape without modification, as both the residual grains and shape modifications would adversely affect the plasmonic properties. Examples of proper antennas for each tested antenna type are shown in Figure 3A. Contrarywise, Figure 3B shows typical examples of antennas evaluated as not proper ones. Due to its smaller surface area, the bar antenna primarily suffers from shape alterations. These occur as a result of uneven sputter rates of various crystallographic orientations of individual grains. Bowtie antennas of larger surface areas are less sensitive to such shape alternations but suffer from residual grains in their vicinity.

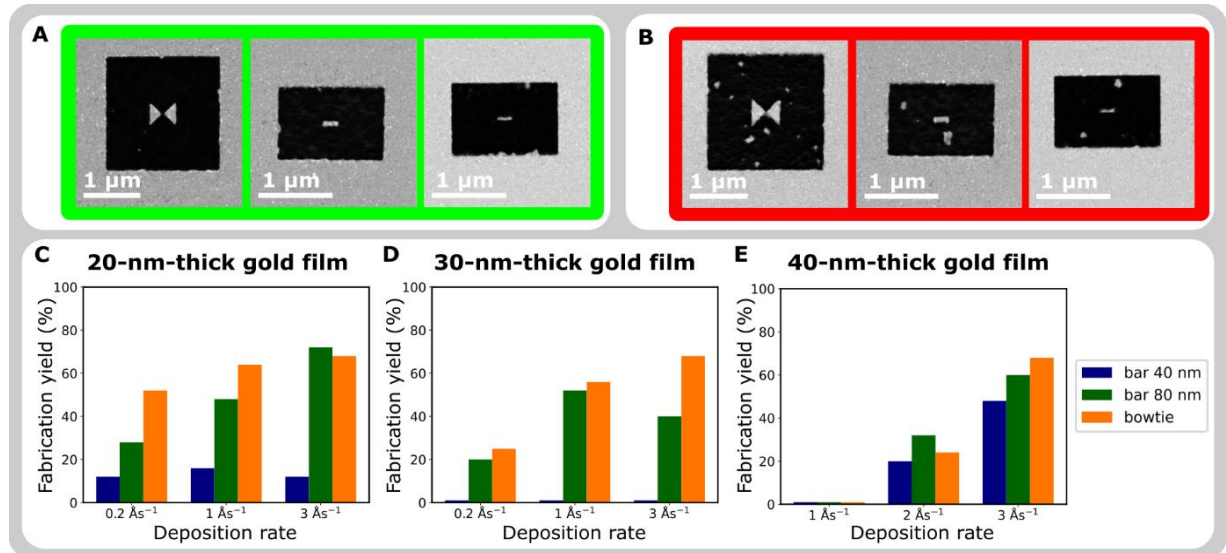

*Figure 3: Antennas fabrication yield. **A**: STEM-HAADF micrographs of proper antennas, i.e., no residual grains are present and the antenna shape is not altered. **B**: STEM-HAADF micrographs of improper antennas, i.e., residual grains in the close vicinity are present and the bow-tie (left) and 40-nm-bar antenna (right) exhibit shape modifications. **C-E**: Fabrication yield for each antenna type as a function of the deposition rate.*

Figure 3C-E shows the fabrication yield for all 27 sets of antennas (3 types, 3 deposition rates, and three film thicknesses). The total number of fabricated antennas in each set was 25. Generally, the lowest fabrication yields have the 40 nm bar antennas as they are the smallest ones which means they are the most sensitive to the quality of the thin films. It is below 20% in the case of 20-nm-thick films (Figure 3C), goes even to zero in the case of 30-nm-thick films (Figure 3D), and reaches higher values for higher deposition rates in the case of 40-nm-thick films (Figure 3E). The increased yield in the thickest films is likely caused by the redeposition of material during the FIB milling. The highest fabrication yield was generally achieved for bow-tie antennas which are the largest ones and therefore seem to be the least sensitive to the quality of the film. The sole exception occurred in the case of the 20-nm-thick film deposited at 3 Ås$^{-1}$, where the highest fabrication yield was achieved for 80 nm bar antennas.

In the case of 20-nm-thick films (Figure 3C), the highest fabrication yield is achieved for the deposition rate of 3 Ås$^{-1}$, moderate for the deposition rate of 1 Ås$^{-1}$, and the lowest one is achieved for the deposition rate of 0.2 Ås$^{-1}$. This means that higher deposition rates are optimal and slow deposition should be avoided. The situation is similar in the case of 30-nm-thick films (Figure 3D). A high fabrication yield is achieved for the deposition rate of 1 Ås$^{-1}$ and 3 Ås$^{-1}$, and the lowest one for the deposition rate of 0.2 Ås$^{-1}$. Again, the higher deposition rates are optimal, and slow depositions should not be carried out. In the case of 40-nm-thick films (Figure 3E), the highest fabrication yield is achieved for the deposition rate of 3 Ås$^{-1}$, a low fabrication yield is achieved for the deposition rate of 2 Ås$^{-1}$, and zero fabrication yield is achieved for the deposition rate of 1 Ås$^{-1}$. Consequently, for a 40-nm-thick film, the optimal deposition rate is 3 Ås$^{-1}$. For 20- and 30-nm-thick films the optimal deposition rate is 1 Ås$^{-1}$ or 3 Ås$^{-1}$.

**Plasmonic properties**

Plasmonic properties of successfully fabricated plasmonic antennas are the final and the most important indicators of the suitability of the respect deposition parameters. This study was performed on sets of bow-tie antennas by STEM-EELS. We focused mostly on two main modes supported by these structures: the transverse dipole (TD) mode and the longitudinal dipole (LD) mode[44]. Figure 4A shows a STEM-ADF micrograph of a bow-tie antenna with a total length of 500 nm. This structure supports the TD mode at 0.8 eV and the LD mode at 1.3 eV. A schematic representation of the TD mode is shown in Figure 4B. The charge oscillates in the direction perpendicular to the antenna's long axis and accumulates in the outer corners of the antenna. The loss probability in EELS is related to the plasmon electric field parallel with the trajectory of the electron beam, which is the largest at the charge antinodes of plasmon oscillations, i.e. at the outer corners of the antenna in the case of the TD mode (Figure 4C). A schematic representation of the LD mode is shown in Figure 4D. The charge oscillates in the direction parallel to the antenna's long axis and accumulates strongly in the gap corners of the antenna. The loss probability then reveals the highest values in the gap of the antenna (Figure 4E). The bow-tie antenna supports higher-order modes, too. For a complete modal analysis, we refer to Ref.[44].

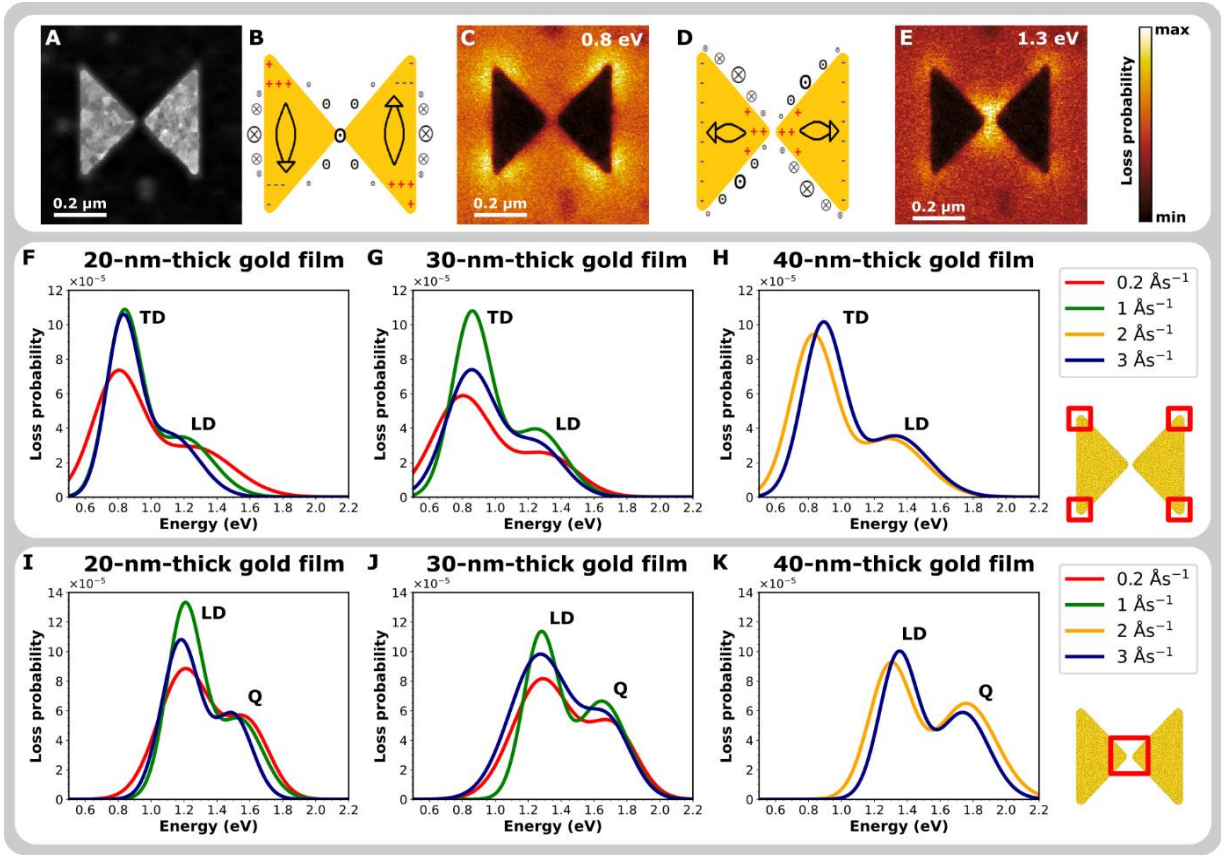

*Figure 4.* Plasmonic properties of bow-tie antennas measured by STEM-EELS. **A:** STEM-ADF micrograph of a bow-tie antenna with a total length of 500 nm. **B:** Schematic representation of a transverse dipole (TD) mode. **C:** Measured loss probability map at the energy of the TD mode (0.8 eV). **D:** Schematic representation of a longitudinal dipole (LD) mode. **E:** Measured loss probability map at the energy of the LD mode (1.3 eV). **F-H:** Averaged fits of loss probability spectra measured at the corners of the bow-tie antennas made of thin films of three different thicknesses and grown at four different deposition rates show capturing the TD and LD mode at the energy of about 0.8 eV and 1.3 eV, respectively. **I-K:** Averaged fits of loss probability spectra measured at the gaps of bow-tie antennas made of thin films of three different thicknesses and grown at four different deposition rates show capturing the LD mode at the energy of about 1.3 eV. Higher-order (Q) modes are noticeable in the energy range 1.5–1.8 eV. The red squares in the insets schematically show the area where the spectra were integrated.

We measured 3 antennas per set. Measured loss probability spectra were integrated over a small region of either the outer corner of the bow-tie (i.e., the electron beam position was around the antenna outer corner) or the bowtie's gap area (i.e., the electron beam position was around the gap). These regions are illustrated in Figure 4 through insets. The experimental spectra were fitted using Gaussians and then averaged to eliminate the influence of minor imperfections in the individual structures. Such processed loss probability spectra are shown in Figure 4F-H for the outer corner positions and in Figure 4I-K for the gap position. The electron beam localized at the antenna outer corner excites both the TD and LD modes (higher-order modes are not visible) while the electron beam situated at the antenna gap excites the LD mode and higher-order modes marked by Q. A closer inspection of the loss probability spectra in Figure 4F-K shows that the plasmon mode energy is not dependent exclusively on the antenna thickness, but also on the deposition rate of the pristine gold layer. The energy shifts due to the deposition rate are up to 0.06 eV. Much pronounced differences are in the peak intensities. In the case of the 20-nm-thick bow-ties, antennas made of the film deposited at 0.2 Ås$^{-1}$ have a significantly lower loss probability than antennas made of the films deposited at 1 Ås$^{-1}$ and 3 Ås$^{-1}$, while the highest loss probability is measured for the antennas made of the film deposited at 1 Ås$^{-1}$ (Figure 4F,I). The situation is rather similar for 30-nm-thick bow-ties. The antennas made of the film deposited at 0.2 Ås$^{-1}$ reach the lowest loss probability, antennas made of the films deposited at 3 Ås$^{-1}$ reach a moderate loss probability, and antennas made of the films deposited at 1 Ås$^{-1}$ reach the highest loss probability (Figure 4G,J). Contrariwise, in the case of 40-nm-thick bow-ties, the antennas made of the films deposited at 2 Ås$^{-1}$ and 3 Ås$^{-1}$ reach both a comparable loss probability (Figure 4H,K).

The maximal loss probability was extracted for TD and LD modes in all evaluated antennas (Figure 5A-C). In the case of 20- and 30-nm-thick bow-ties, the highest values for both TD and LD modes are reached by antennas fabricated from the films deposited at 1 Ås$^{-1}$ (Figure 5A,B). In the case of 40-nm-thick bow-ties, the highest values for both TD and LD modes are reached by antennas fabricated from the films deposited at 3 Ås$^{-1}$ (Figure 5C).

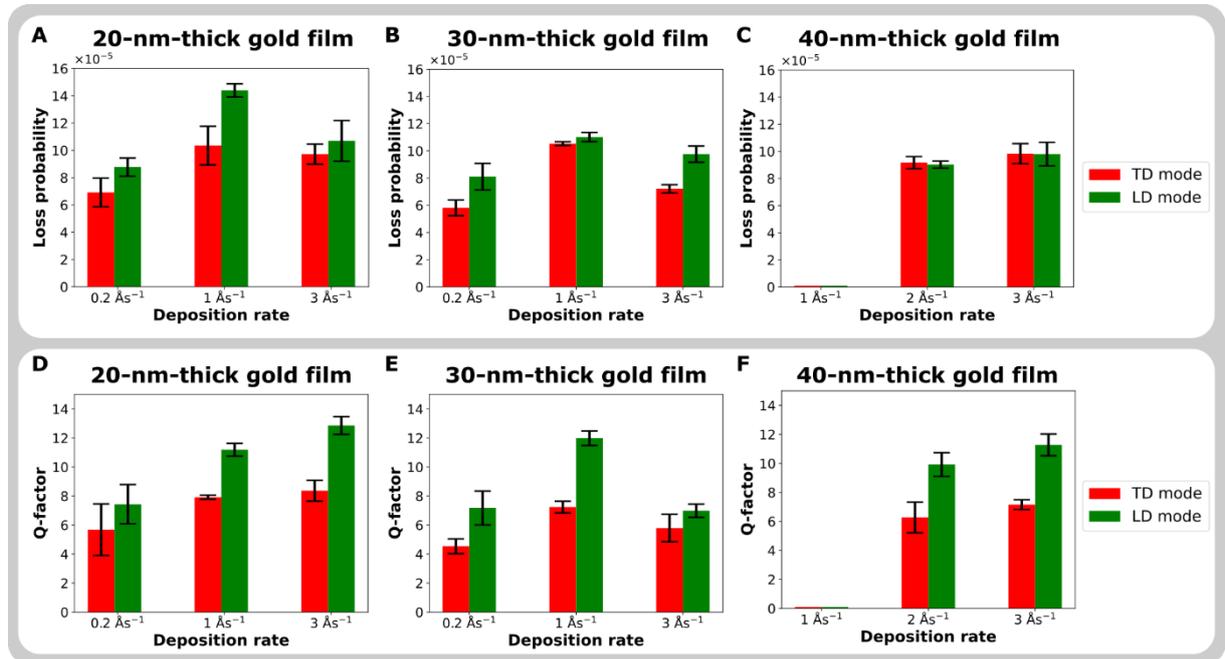

*Figure 5.* Properties of the two main modes of bow-tie antennas for the three film thicknesses as a function of the deposition rate. ***A-C:*** *Loss probability peak maxima of transversal dipole (TD) and longitudinal dipole (LD) modes* ***D-F:*** *Q-factors of TD and LD modes. Note that no successful antennas were achieved in the case of 40-nm-thick films deposited at 1 Ås$^{-1}$.*

Finally, the Q-factors, defined as the LSPR energy divided by its FWHM, were evaluated for TD and LD modes in all antennas (Figure 5D-F). In all cases, the Q-factors of LD modes are higher than those of TM modes. In the case of 20-nm-thick bow-ties, the lowest Q-factors are reached by antennas fabricated from the films deposited at 0.2 Ås$^{-1}$ and read 5.7 ± 1.8 for the TD mode and 7.4 ± 1.4 for the LD mode (Figure 5D). Higher Q-factors are reached for both TD and LD modes by antennas fabricated from the films deposited at 1 Ås$^{-1}$ and 3 Ås$^{-1}$ and read 7.9 ± 0.1 and 8.4 ± 0.8 for the TD mode, and 11.2 ± 0.4 and 12.9 ± 0.6 for the LD mode, respectively. In the case of 30-nm-thick bow-ties, the highest Q-factors are reached by antennas fabricated from the films deposited at 1 Ås$^{-1}$ and read 7.2 ± 0.4 for the TD mode and 12.0 ± 0.5 for the LD mode (Figure 5E). Considerably lower Q-factors are reached for both modes by antennas fabricated from the films deposited at 0.2 Ås$^{-1}$ and 3 Ås$^{-1}$ and read 4.5 ± 0.5 and 5.8 ± 0.9 for the TD mode, and 7.2 ± 1.2 and 7.0 ± 0.5 for the LD mode, respectively. In the case of 40-nm-thick bow-ties, slightly higher Q-factors for both modes are reached by antennas fabricated from the films deposited at 3 Ås$^{-1}$ than by antennas fabricated from the films deposited at 2 Ås$^{-1}$ (Figure 5F) and read 7.2 ± 0.3 over 6.3 ± 1.1 for the TD mode, and 11.7 ± 0.8 over 9.9 ± 0.8. for the LD mode, respectively.

Consequently, the best plasmonic properties have the bow-ties fabricated from the 20- and 30-nm thick films deposited at the deposition rate of 1 Ås$^{-1}$. In the case of 40-nm-thick films, the best plasmonic properties have the bow-ties fabricated from the film deposited at the deposition rate of 3 Ås$^{-1}$.

**Conclusion**

In summary, we have deposited gold films of three different thicknesses at various deposition rates and evaluated their morphology and crystallography, the ease of fabricating plasmonic antennas using FIB lithography, and their plasmonic properties. The films are homogeneous with the surface roughness increasing with the deposition rate. The dominant crystallographic orientation perpendicular to the surface is (111) in all films. Consequently, the films deposited at slower deposition rates seemed to be ideal candidates for use in plasmonics based on the structural properties of the films. However, antennas fabricated using FIB lithography experienced shape alterations due to uneven removal of gold caused by different crystallographic orientations. Interestingly, the thickness of the film did not seem to have a significant effect on the ease of antenna fabrication or the yield of successful antennas. In some cases, it has been observed that the redeposition of material during FIB lithography can assist in the repair of uneven edges of antennas, which can lead to the easier fabrication of very thin antennas. The highest fabrication yield was achieved for films deposited at the highest deposition rate of 3 $\text{Ås}^{-1}$ and the lowest for films deposited at the slowest deposition rate of 0.2 $\text{Ås}^{-1}$.

Finally, the most important criterion is the best plasmonic behavior. The best plasmonic properties have the bow-ties fabricated from the 20- and 30-nm thick films deposited at the deposition rate of 1 $\text{Ås}^{-1}$ and the bow-ties fabricated from the 40-nm thick films deposited at the deposition rate of 3 $\text{Ås}^{-1}$. The loss probability, and subsequently the plasmon resonance intensity, were the highest for these films. Additionally, the Q-factors were among the highest. Once again, it appears that there is no conclusive evidence to suggest that increasing the film thickness harms the plasmonic properties of the fabricated antennas. It is worth noting that the antennas made of films deposited at 0.2 $\text{Ås}^{-1}$ performed the worst, exhibiting smaller loss probability maxima, broader peaks, and thus lower Q-factors. To conclude, in our deposition method, the optimal gold thin film for plasmonic antenna fabrication with a thickness of 20 nm and 30 nm should be deposited at the deposition rate of around 1 $\text{Ås}^{-1}$. The thicker 40 nm film should be deposited at a higher deposition rate like 3 $\text{Ås}^{-1}$.

**Acknowledgment**

This research was supported by project Quantum materials for applications in sustainable technologies (QM4ST, project No. CZ.02.01.01/00/22_008/0004572, P JAK, call Excellent Research), project CzechNanoLab (project No. LM2023051, MEYS CR), and by Brno University of Technology (project No. FSI-S-23-8336).